\documentclass[twocolumn]{aastex6}
\usepackage{amsmath,amssymb}
\usepackage{hyperref}
\newcommand{\itu}{1}
\newcommand{\fg}{2}
\newcommand{\iu}{3}
\newcommand{\harvard}{4}

\begin{document}

\title[Inclination angle and braking index evolution of pulsars with plasma-filled magnetosphere]{Inclination angle and braking index evolution of pulsars with plasma-filled magnetosphere: application to high braking index of PSR~J1640--4631}


\author{K.~Y. Ek\c{s}i\altaffilmark{\itu,\fg},
	   I.~C. Anda{\c c}\altaffilmark{\itu},
	   S. {\c C}{\i}k{\i}nto{\u g}lu\altaffilmark{\itu},
	   E. G\"ugercino{\u g}lu\altaffilmark{\iu},
	   A. Vahdat Motlagh\altaffilmark{\itu},
	   and B. K{\i}z{\i}ltan\altaffilmark{\harvard}}


\affil{\altaffilmark{\itu} Istanbul Technical University,  Faculty  of Science  and  Letters,  Physics Engineering  Department,
  34469,  Istanbul, Turkey}
\affil{\altaffilmark{\fg}  Feza G{\"u}rsey Center for Physics and Mathematics, Bo{\u g}azi{\c c}i University, 34684, {\c C}engelk{\" o}y, Istanbul, Turkey}
	\affil{\altaffilmark{\iu} Istanbul University, Faculty of Science, Department of Astronomy and Space Sciences, Beyaz{\i}t, 34119, Istanbul, Turkey}
	\affil{\altaffilmark{\harvard} Harvard-Smithsonian Center for Astrophysics,  60 Garden Street, Cambridge, MA 02138}



\begin{abstract}
The recently discovered rotationally powered pulsar PSR~J1640--4631 is the first to have a braking index measured, with high enough precision, that is greater than three. 
An inclined magnetic rotator in vacuum or plasma would be subject not only to spin-down but also to an alignment torque. The vacuum model can address the braking index only for an almost orthogonal rotator that is incompatible with the single peaked pulse profile.
The magnetic dipole model with the corotating plasma predicts braking indices between $3-3.25$. We find that the braking index of $3.15$ is consistent with two different inclination angles, 
$18.5\pm 3$ degrees and $56 \pm 4$ degrees. The smaller angle is preferred given the pulse profile has a single peak and the radio output of the source is weak. 
We infer the change in the inclination angle to be at the rate $-0.23$ degrees per century, three times smaller in absolute value than the rate recently observed for the Crab pulsar.
\end{abstract}


\section{Introduction}
\label{sec:intro}

The rotationally powered pulsar PSR~J1640--4631 recently discovered by NuSTAR in X-rays may be holding a clue about the origins of pulsar spin down. PSR~J1640--4631 has a spin frequency of $\nu = 4.843$~Hz and a spin-down rate of 
$\dot{\nu} = - 2.28 \times 10^{-11}$~s$^{-2}$  \citep{arc+16} implying a  characteristic age of $\tau_{\rm c} \equiv -\nu /2\dot{\nu} = 3370$~years and spin-down power of $5.5\times 10^{36}$~erg~s$^{-1}$.
Interestingly, it is the first pulsar to have a braking index (i.e., $n \equiv \nu \ddot{\nu}/\dot{\nu}^2$)
greater than three, $n=3.15 \pm 0.03$ \citep{arc+16},  measured with high precision. 
This is unusual given the previous measurements \citep{lyn+93,liv+07,lyn+96,esp+11,liv11,roy+12,ant+15,arc+15,fer+15} for 8 other pulsars with braking indices $n<3$  \citep[see][for a compilation]{lyn+15}.

A dipole rotating in vacuum is subject to magnetic dipole radiation (MDR) torque with spin-down \textit{and} alignment components respectively given as
\begin{align}
I \frac{d \Omega}{dt} &= -\frac{2 \mu^2 \Omega^3}{3c^3} \sin^2 \alpha \label{mdr1}  \\
I \frac{d \alpha}{dt} &= -\frac{2 \mu^2 \Omega^2}{3c^3} \sin \alpha \cos \alpha \label{mdr2}
\end{align}
\citep{mic70,dav70} where $\Omega=2\pi \nu$ is the angular velocity and  $\mu$ is the magnetic moment 
of the star, $\alpha$ is the inclination angle between rotation and magnetic axis and $c$ is the speed of light. 
This model predicts a braking index of 
\begin{equation}
n=3 + 2\cot^2 \alpha
\label{n_vacuum}
\end{equation} 
\citep{mic70,dav70} which is always greater than 3 and diverges for small inclination angles.
Such a model predicts rapid alignment within a spin-down time-scale. According to this vacuum magnetic dipole model a pulsar would stop spinning down when alignment is achieved in obvious contradiction with observations. It was suggested by  \citet{gol70a} that the progress of alignment would be slowed down by dissipative processes for a non-spherical pulsar. 
Therefore, the effect of the alignment torque given in \autoref{mdr2} has not been adequately appreciated in the literature. As a result it has been rarely used together with the spin-down torque given in \autoref{mdr1} although the alignment torque is an intrinsic component of the torque due to the magnetic dipole.

The recent observational evidence suggesting an increasing inclination angle of the Crab pulsar \citep{lyn+13} (see also \citet{ge+16}) implies that the magnetospheric torques can dominate the dissipative processes invoked by \citet{gol70a}. Yet the increasing inclination angle of Crab pulsar may require a further ingredient such as the presence of return currents in the 
magnetosphere \citep{bes07} or precession \citep{arz+15,zan15}. But an orthogonal rotator with plasma-filled force-free magnetosphere \citep{phi+14} requires a much larger current than used in \citet{bes07} so that the minimum spin-down energy losses correspond to that of an aligned rotator in the end. There is also statistical evidence that the inclination angle of pulsars tend to achieve alignment in the long term \citep{you+10,lyn88}.

The pulsar is unlikely to be rotating in vacuum; its magnetosphere is expected to be filled by a corotating plasma formed through charged particles ripped off from the surface of the neutron star \citep{gol69} and thereafter accelerated by rotation induced electric field along curved magnetic field lines to give an excess of electron--positron pair discharges \citep{philippov14}. As shown by recent simulations \citep{chen14,philippov15} only such a configuration is capable of maintaining a pulsar active. The spin-down \citep{spi06} and alignment \citep{phi+14} torques in the presence of a corotating plasma also predict alignment, but at a slower pace compared to that of the vacuum model. Moreover, the model predicts a braking index of $n=3-3.25$  depending on the inclination angle \citep{arz+15}. This obviates any necessity to invoke other assumptions to address the measured braking index of  PSR~J1640--4631, $n=3.15 \pm 0.03$ \citep{arc+16}.

We show within the framework of the plasma-filled magnetosphere model that there are two possible solutions for the inclination angle of  PSR~J1640--4631. We infer the magnetic dipole moment and rate of change of the inclination angle of the pulsar, and provide the implications on the evolution of pulsars on the $P-\dot{P}$-diagram.
In \S~\ref{sec:brake} we describe the model details and its consequences. In \S~\ref{sec:discuss} we discuss the results in view of less-than-three braking indices.

\section{Evolution of pulsars with plasma filled magnetospheres}
\label{sec:brake}

The spin-down \citep{spi06} and alignment \citep{phi+14} torques in the presence of a corotating plasma are
 \begin{align}
I \frac{d \Omega}{dt} &= -\frac{ \mu^2 \Omega^3}{c^3} (1 + \sin^2 \alpha ) \label{mdr3}  \\
I \frac{d \alpha}{dt} &= -\frac{ \mu^2 \Omega^2}{c^3} \sin \alpha \cos \alpha \label{mdr4}.
\end{align}
Although these equations seem to be only slightly different than the equations of the vacuum model given in \autoref{mdr1} and \autoref{mdr2}, they predict a slower alignment \citep{phi+14} and imply the spin-down of the pulsar even when alignment is achieved as originally predicted by \citet{gol69}.

\begin{figure}
\centering
\includegraphics[width=\columnwidth]{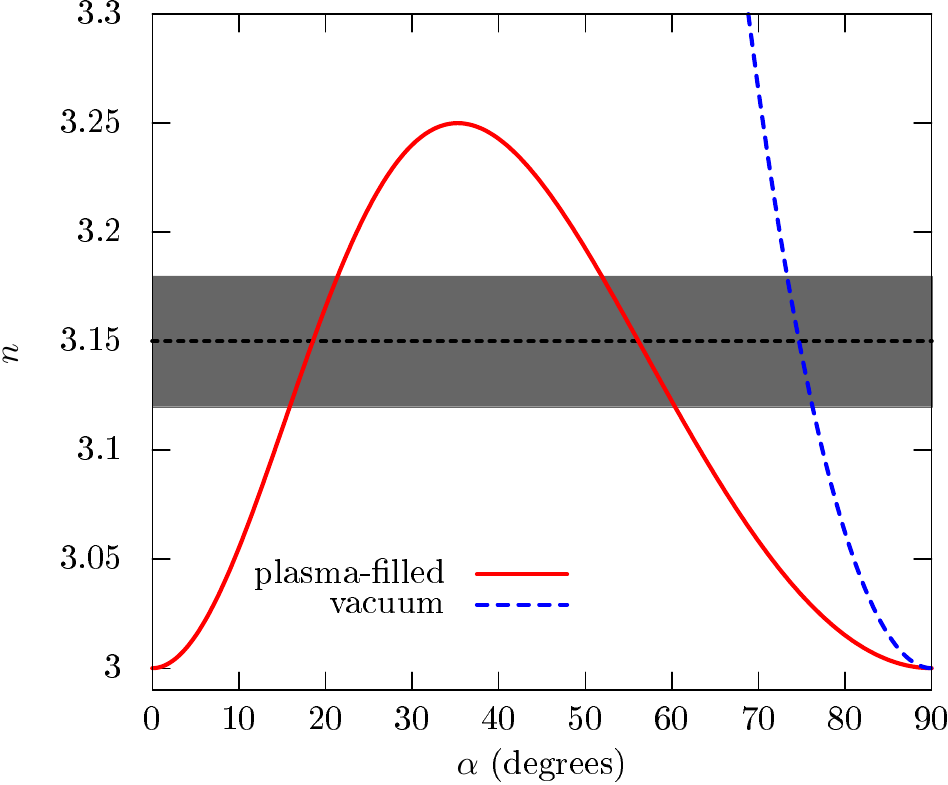}
\caption{The braking index versus the inclination angle for the pulsars. The solid line (red in the electronic version) stands for the prediction of the plasma filled magnetosphere model given in \autoref{n_2}; the long-dashed line (blue in the electronic version) stands for the prediction of the vacuum model given in \autoref{n_vacuum}. The horizontal short-dashed line (black) and the surrounding shaded region denote, respectively, the measured braking index of $n=3.15$ and its possible range given by the uncertainty $0.03$ for the  PSR~J1640--4631.}
\label{fig:brake}
\end{figure}

Using  only \autoref{mdr3}, the braking index in this model is 
\begin{equation}
n = 3 - 4\dot{\alpha} \tau_{\rm c} u
\label{n_1}
\end{equation}
where
\begin{equation}
u(\alpha) \equiv \frac{\sin \alpha \cos \alpha}{1 + \sin^2 \alpha}.
\label{u_alpha}
\end{equation}
Using \autoref{mdr4} in \autoref{n_1} the braking index can be calculated as
\begin{equation}
n = 3 + 2u^2
\label{n_2}
\end{equation}
\citep{arz+15} which implies $3<n<3.25$ depending on the inclination angle and does not diverge as its vacuum counterpart for small inclination angles.
In \autoref{fig:brake} we show this prediction of the plasma-filled magnetosphere model \citep{spi06,phi+14} together with that of the vacuum model \citep{mic70,dav70} given in \autoref{n_vacuum}. Accordingly, the plasma-filled model can explain the braking index of $3.15\pm 0.03$ observed from PSR~J1640--4631 \citep{arc+16} for two different inclination angles, $18.5\pm 3$ degrees and $56 \pm 4$ degrees. Of these we favour the smaller value as the pulse profile shows a single peak \citep{arc+16}.

\begin{figure*}
\centering
\includegraphics[width=1.6\columnwidth]{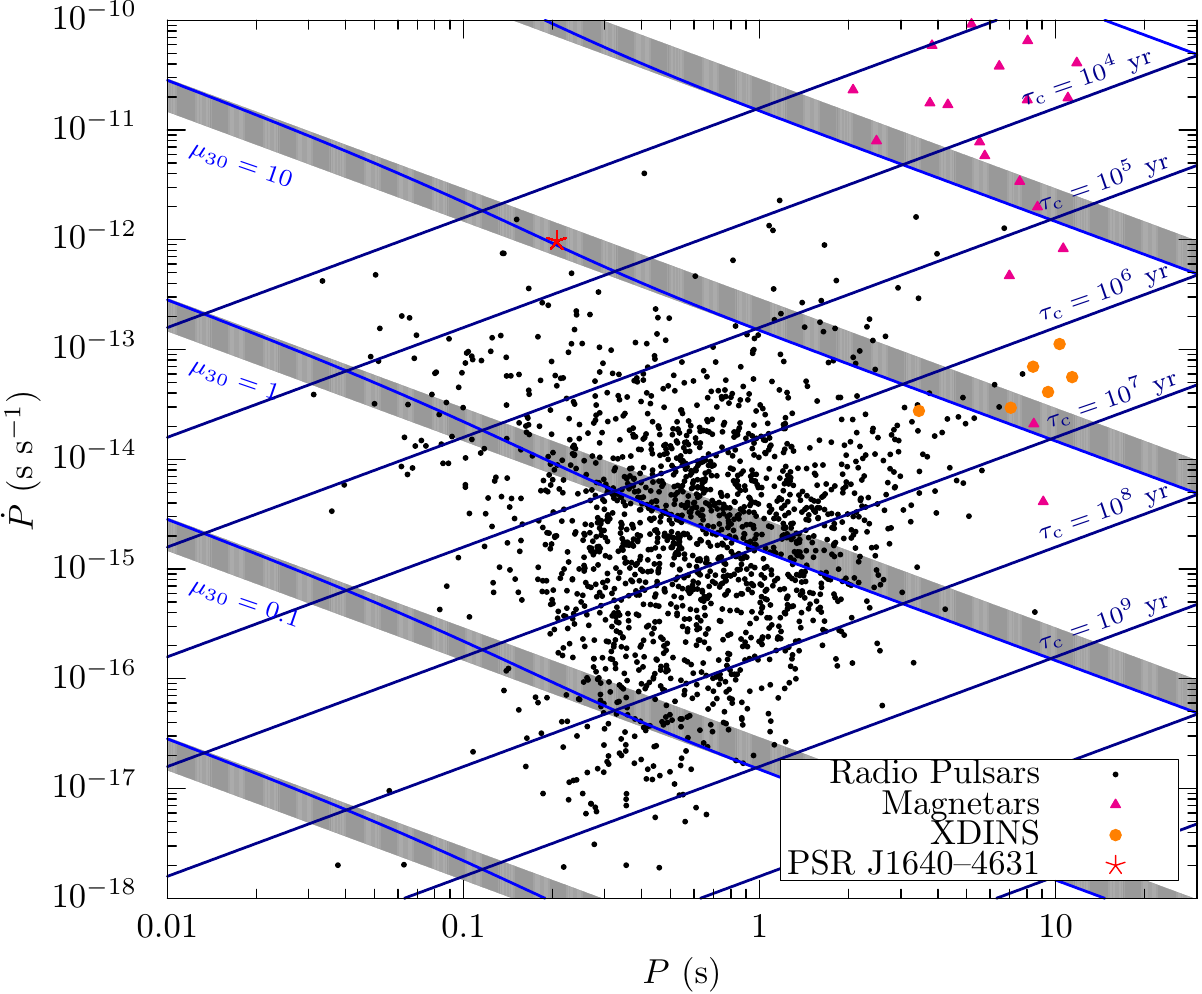}
\caption{The pulsar PSR~J1640--4631 on the $P-\dot{P}$ diagram.  The rotationally powered pulsars are shown with small points. Magnetars are shown with triangles (red in the electronic version), X-ray dim neutron stars (XDINS) are shown with solid circles (orange in the electronic version) and the pulsar PSR~J1640--4631 is shown with the star symbol (red in the electronic version). The equi-magnetic field lines  are determined according to the model presented in this section. The shaded bands involving each equi-magnetic field line are the envelopes of them for the full inclination angle range. Equi-characteristic age lines are also shown (dark blue in the electronic version). The data is taken from ATNF pulsar catalogue at URL \url{http://www.atnf.csiro.au/people/pulsar/psrcat/} \citep{man+05}.}
\label{fig:p-pdot}
\end{figure*}

From the measured braking index of  PSR~J1640--4631 \citep{arc+16} by using \autoref{n_2} we find $u=\sqrt{(n-3)/2}=0.274 \pm 0.025$. Using this result in  \autoref{n_1} the rate of decrease of the inclination angle can be found as
\begin{equation}
\dot{\alpha} = -(0.23\pm 0.05)^\circ ~{\rm century}^{-1}.
\label{rate} 
\end{equation}
This value is about three times smaller (in absolute value) than the measured increasing inclination angle of the Crab pulsar \citep{lyn+13}.

By dividing \autoref{mdr3} with 
\autoref{mdr4} and upon integration one finds $ \cos^2 \alpha /  P \sin \alpha$ is a constant throughout the evolution of the pulsar \citep{phi+14}; as $P=2\pi/\Omega$ increases $\cos^2 \alpha / \sin \alpha$ should also increase which is achieved for small values of $\alpha$. Thus we define
\begin{equation}
A \equiv \frac{ \cos^2 \alpha_0}{P_0 \sin \alpha_0 } = \frac{ \cos^2 \alpha}{P \sin \alpha }
\label{omega_alpha}
\end{equation}
where  $P_0$ is the initial spin period and $\alpha_0$ is the initial inclination angle. For $P=0.2065$~s and $\alpha=18.5^\circ$ we find $A=13.72$ for this pulsar. Solving $\sin \alpha$ from the above equation gives
\begin{equation}
\sin \alpha = \sqrt{1 + \left(  \frac{AP}{2} \right)^2 } - \frac{AP}{2}.
\label{sina}
\end{equation}
Accordingly the period derivative is related to the period as 
\begin{equation}
\dot{P} =\frac{(2\pi)^2 \mu^2}{I c^3 P}\left( 1+\sin^{2}\alpha \right)
\label{pdot-p}
\end{equation}
where $\sin \alpha$ is from \autoref{sina} and $A$ is from \autoref{omega_alpha}.
We infer the magnetic dipole moment of the pulsar in units of $10^{30}$~G~cm$^3$ as $\mu_{30} = 11.2$  from the measured spin frequency and its derivative from \autoref{mdr3} employing $\alpha = 18.5^\circ$ and $I=10^{45}$~g~cm$^2$. The position of the pulsar is shown in the $P-\dot{P}$-diagram in \autoref{fig:p-pdot}.

\section{Discussion}
\label{sec:discuss}

We have shown that the greater-than-three braking index of the recently discovered pulsar PSR~J1640--4631 \citep{arc+16} is readily explained by the plasma-filled magnetic dipole model \citep{gol69,spi06,phi+14,arz+15}. The only free parameter of the model is the inclination angle and there are two possible solutions; $18.5\pm 3$ degrees and $56 \pm 4$ degrees. 
The smaller inclination angle is favoured because of the single peak in the pulse profile \citep{ran83,ran90,wel08,han10}. The relatively small radio output of this object, 
with an upper limit of 0.018 mJy at 1.4 GHz \citep{arc+16} may also be a consequence of this small inclination angle of the pulsar.

Note that the vacuum model as employed by \citet{arc+16} and also seen in \autoref{fig:brake} predicts an almost orthogonal rotator. As this is not compatible with the single peaked pulse profile, \citet{arc+16} disfavoured the idea that the alignment of the rotation and magnetic axes, i.e.\ decrement in the inclination angle, is the cause of the greater-than-three braking index of PSR~J1640--4631 and suggested that the quadrupole magnetic moment of this object could be important. The model with plasma-filled magnetosphere, on the other hand does not require an orthogonal rotator to produce the observed braking index of PSR~J1640--4631 as seen in \autoref{fig:brake}.
Given that the presence of the charged particles around pulsars is well established  \citep[see e.g.][]{gol69} we conclude that the evolution of the inclination angle of PSR~J1640--4631 towards an aligned rotator, within the framework of plasma-filled magnetosphere model \citep{spi06,phi+14}, is a simpler and better understood explanation of the observed greater-than-three braking index. 

Another possibility for explaining the greater-than-three braking index could be an anomalous $\ddot{\nu}$ resulting from a glitch \citep{alp06}. We note that $n=3.15\pm 0.03 $ is not an anomalous braking index within the framework of the plasma-filled magnetosphere model \citep{arz+15}.

\citet{arc+16} favours the quadrupole structure of the magnetosphere \citep{pet15} for explaining the greater-than-three braking index of PSR~J1640--4631. We show here that this process alone, with no assistance from the alignment torque, would require very strong quadrupole fields to increase $n$ up to $3.15$. In this case the spin-down luminosity due to the quadrupole field has to be $(n-3)/(5-n) \simeq 10 \%$ of the dipole spin-down luminosity. This implies the Poynting flux of the quadrupole field to be $10 \%$ of that of the dipole field. This then requires the quadrupole field  to be $\sqrt{0.1}\simeq 0.3$ times the dipole field at the light cylinder, $R_{\rm L} = c/\Omega \simeq 10^9~{\rm cm}$. Assuming a stellar radius of $R_{\ast} = 10^6~{\rm cm}$ this implies a quadrupole field $\sim 300$ times stronger than the dipole field! Given that the dipole field at the equator is $B_{\rm d} = \mu/R^3 \simeq 10^{13}~{\rm G}$, we infer a quadrupole field of $B_{\rm q} \simeq 3 \times 10^{15}~{\rm G}$ for this object. The radio polarimetry measurements, for normal pulsars, indicate that the field is of dipolar form at heights $ \lesssim 30 R_{\ast}$ where radio emission is generated \citep{kar07}. Hence such a discrepancy between the dipole and quadrupole fields at the surface is unusual for normal pulsars. 
Yet we can not exclude the possibility  that PSR~J1640--4631 has super-strong quadrupole fields unlike that of ordinary pulsars. Presence of such super-strong quadrupole fields would render the object similar to the  $\sim 800$ year old pulsar PSR J1846-0258 located in Kesteven 75 supernova remnant \citep{got+00} which emitted several magnetar-like bursts \citep{gav+08}. The dipole field of PSR J1846-0258 is about $5\times 10^{13}~{\rm G}$ i.e.\ 5 times stronger than that of PSR~J1640--4631, and is marginally beyond the quantum-critical limit, $B_{\rm q} = 4.4\times 10^{13}~{\rm G}$. If PSR~J1640--4631 also has such strong quadrupole fields near the surface it may be expected to show magnetar-like bursts likely associated with some glitch activity in the future. Such pulsars with super-strong quadrupole fields may be another manifestation of low-B magnetars \citep{rea+10} as argued by \citet{per11}.

Given the measured greater-than-three braking index of PSR~J1640--4631 can be addressed by employing the conventional view of a pulsar with corotating magnetosphere, the question naturally arises why most other measured braking indices are less than three. Many different models have been suggested to explain the less-than-three braking indices: Some models invoke an external torque similar to that from stellar winds \citep[e.g.][]{ou+16} or from a putative fallback disk \citep[e.g.][]{mic81,men+01,ozs+14}. Some others modify the magnetic dipole model by suggesting that the magnetic dipole fields are increasing \citep{bla88} e.g.\ due to diffusion ensuing post-supernova field burial by accreted matter \citep[e.g.][]{gep+99,esp+11,gun13,ho15} or by poloidal field growth at the expense of interior toroidal field decay through Hall effect \citep{gou15}. Another interesting suggestion is the increasing inclination angles of pulsars are increasing in the long term \citep{bes+84} as possibly observed in the Crab pulsar \citep{lyn+13}. Yet the latter observation might  result from precession \citep{zan15,arz+15}. Yet another modification of the dipole model emphasizes the finite size effects for the dipole given the presence of a corotating plasma \citep{mel97}.

We showed that a magnetic dipole within a plasma filled magnetosphere does produce greater-than-three braking indices. Therefore, invoking a combination of mechanisms (e.g.\ wind and magnetic quadrupole moments assisting the plasma filled magnetosphere) appears unnecessary to explain the braking index of PSR~J1640--4631. It is particularly unclear how these physical mechanism affect the inclination angle.

Any process that could potentially account for the less- than-three braking index should start reducing the the braking index from higher values (i.e.\ $n = 3 - 3.25$ as predicted by \citet{arz+15}) rather than the canonical value $n=3$ assumed by neglecting the alignment torque.

\section*{Acknowledgements}

KYE and BK acknowledge the Scientific and Technological Council of TURKEY 
(TUBITAK) for BIDEB 2221 visiting scholar award and E.G. for support under the grant
113F354. We thank the referee, Alexander Philippov, for insightful comments that improved the paper.



\end{document}